\newcommand{\clcl}{\emph{cluster-cluster }}

\documentclass{PoS}

\title{Revisited ``Cluster-Cluster'' VLBI with future 
       multi-beam low frequency radio interferometers }

\ShortTitle{``Cluster-Cluster'' VLBI with future radio interferometers}

\author{\speaker{Mar\'{\i}a J. Rioja}
\thanks{On secondment from OAN, Spain.}, 
        and Richard Dodson \\
        ICRAR, UWA, Perth, Australia\\
        E-mail: \email{maria.rioja@uwa.edu.au},
                \email{richard.dodson@uwa.edu.au}}

\author{Richard W. Porcas \\
        Max-Planck-Institut f\"ur Radioastronomie, Bonn, Germany \\
        E-mail: \email{porcas@mpifr-bonn.mpg.de}}

\author{Dick Ferris and John Reynolds \\
        ATNF, CSIRO, Sydney, Australia \\
        E-mail: \email{Dick.Ferris@atnf.csiro.au},
                \email{John.Reynolds@csiro.au}}

\author{Tetsuo Sasao \\ 
        E-mail: \email{sasaotts@nifty.com}}

\author{Richard T. Schilizzi \\
        SKA Program Development Office, The University of Manchester, UK \\
        E-mail: \email{schilizzi@skatelescope.org}}

              \abstract{We revisit the {\it ``Cluster-Cluster''} or
                multi-view VLBI technique from the perspective of its
                syner\-gy with the multi-beam features inherent in the
                Australian Square Kilometer Array Pathfinder (ASKAP)
                and its potential to improve the outcomes of VLBI
                observations with ASKAP. We include a list of
                candidate VLBI sites that already support or can be
                upgraded to support multi-view VLBI located in
                Australia and overseas, and which have common visibility
                with ASKAP.  The results of our previous \clcl
                observations at 1.6 GHz demonstrated the advantages of
                this configuration to calibrate the ionospheric
                distortions responsible for the loss of positional
                accuracy at low frequencies, using multiple
                calibrators in a range between 1 to 6 degrees away
                from the target.  Therefore, we conclude that joint
                observations of ASKAP with other multi-view sites
                using \clcl techniques would improve the outcomes of
                the high spatial resolution component of ASKAP applied to
                astrometric projects, achieving higher precision
                for many more targets, and with
                lower detection thresholds. Also very wide-field VLBI
                mapping becomes a possibility.  Looking to the future,
                this would contribute to the development of new
                techniques that are relevant for future high
                resolution observations with the SKA.}
\FullConference{Science and Technology of Long Baseline Real-Time 
Interferometry: \\
The 8th International e-VLBI Workshop, EXPReS09\\
		 June 22 - 26 2009\\
		 Madrid, Spain}

\begin{document}

%Logical points:
%- Conventional phase ref very useful
%- Ionospheric propagation effects limit the application of
%  conventional phase referencing techniques to VLBI observations at low
%  frequencies, either for astrometry or increased sensitivity (coherence time).
%- The unknown spatial structure of the ionosphere being the main
%reason - prevents direct transfering of calibration terms between
%observations of sources along different directions.
%- Also, the ionosphere has temporal structure.
%- Strategies used: The use multiple (2) calibrators, antennas 
%  switching/succesively observed, can mitigate the effect of the
%  spatial structure. ``In beam'' phase referencing - beam within ASKAP 
%  small tied array beam 
%- Use cluster-cluster: high sensitivity tieding multiple antennas,
%  spatial and temporal compensation, 
%- Next generation instruments for observations at low frequencies  have
%  inherent multi-beam nature, mention wide field of view,
%  and will have a high resolution component.

%HYPOTHESIS
%The application of Multi-view techniques to the next
%generation of multi-beam instruments at low frequencies holds a great 
%potential for/achieve high precision astrometry with VLBI at low frequencies.  

\section {Introduction}

%Conventional phase referencing techniques achieve the highest 
%precision relative astrometric measurements, and lower the threshold
%of detection in VLBI observations of a target source, with alternating
%observations of a nearby calibrator source,  
%at frequencies between a few and tens of GHz.

Conventional Very Long Baseline Interferometry (VLBI) 
phase-referencing techniques can achieve the highest 
precision relative astrometric measurements
%and lower the threshold of detection in VLBI observations 
of a target source, using alternating observations of a nearby
calibrator source,  at frequencies between a few and tens of GHz.
%(interpolated solutions between consecutive observations of the
%calibrator) 
%(switching observations between the target and calibrator 
%sources), 
%(gives VLBI the capability to achieve) 
%the highest precision astrometric measurements, and enhances the 
%sensitivity of the array for detection of weak sources. \\
At lower frequencies, i.e., below 2 GHz, its application is constrained.
%(exponencially??the lower the freq the more the limitation) 
%There are however a number of limitations on the application of
%conventional phase referencing to lower frequency regimes 
%as observed with ASKAP. 
The cause of this limitation is the unpredictability of both the
temporal variations and the spatial irregularities in the plasma density
in the ionosphere, which introduce differential path variations between the
observations of the two sources, even for very fast switching times and 
small source separations.
%temporal fluctuations and 
%direction-dependent shifts between the observations of the two sources,
%(between/in the observations of the two sources),
%respectively, 
%The difference, for/of/between? the two sources, in the ionospheric
%delays may be large, 
%(atmospheric phase corruptions do not fully cancel.)  
%Having a calibrator within the antenna beam size of the target source
%would provide an ideal configuration, sufficiently nearby  calibrator 
%In general this is not possible and
The temporal and spatial differential ionospheric fluctuations 
degrade the positional accuracy  of the technique and, eventually, 
prevent the phase connection process and the use of conventional phase 
referencing. These fluctuations are larger at lower frequencies.\\

In general, observations which involve more than one calibrator
have demonstrated advantages for astrometric VLBI at low frequencies. 
%source provides a superior
%ionospheric calibration than a single nodding calibrator
%strategies for astrometry at low frequencies involve
The exception is the ideal but unusual configuration when 
%Leaving aside the unlikely and ideal configuration of having
%simultaneous observations of a target and a strong calibrator that 
%lie within the VLBI antenna beam (termed "in-beam" calibrator)
a target and a strong calibrator lie within the field-of-view (FoV) of the 
VLBI antennas  (an ``in-beam'' calibrator), and thus can be observed
simultaneously.
% (Rioja \etal XX).
%Next follows a few examples of configurations that 
%have shown advantages for
%% (ionospheric compensation in) 
%astrometric VLBI at low frequencies. 
%provides a superior
%ionospheric calibration than a single nodding calibrator
A useful variation of this 
%mentioned above 
combines the observations of an ``in-beam'' weak calibrator
source and nodding to a more distant strong calibrator.  
%, hence it is called ``in-beam'' calibrator. 
The observations of the strong calibrator are used to remove 
the first-order atmospheric effects; then the observations of 
the weak source, which is observed along with the target source, 
provide further adjustments (a fine tuning) of the spatial 
and temporal fluctuations, with longer coherence times \cite{fom02}.
%(Fomalont \etal \,2002).
%temporal and spatial differential residuals with longer coherence times.
%the observations of a target source 
%In-beam observations very promising
%The in beam advantages (no temporal interpolation) can be extended 
%beyond the severe constraints
%for a suitable calibrator, and using a relatively weak source ``in
%beam'' supported by a nearby source. 
The results obtained with this approach are promising; however
its widespread application
% of this approach is not widespread because it is limited by
% sensitivity (OR however the widespread/general application of this
% approach
is still limited by sensitivity.  
%%%%REFERENCE (E.B. Fomalont, W.M.Goss, A.J.Beasley, S.Chatterjee, AJ, 117, 3025-3030, 1999)
%%%%%%%
%Succesfully applied, Very promising, but its general/wide/global application
%is sensitivity limited. Restricted application... to targets with...
%Still, both configurations very demanding on calibrator positions -
%are not applicable to every source. 
Another useful approach is when there are two calibrators aligned
with, but on opposite sides of, the target. 
%provides a superior
%ionospheric calibration than a single nodding calibrator
%(source interpolation) 
During the observations the telescopes alternate
every few minutes between the three sources, and in the analysis
successive scans on the calibrators are used for the spatial and
temporal interpolation to the enclosed position and scan time
of the target source \cite{fom99}.
%(Fomalont \etal \,1999). 
%%%%%REFERENCE (Fomalont, 2002, Proceedings of
%the 6th European VLBI Network Symposium, Eds. Ros, E., Porcas, R.W.,
%Lobanov, A.P., and Zensus, J.A.). 
The unlikely source configuration
required for this approach to work results in limited applicability,
and the overhead calibration time is large. \\

This contribution is concerned with the multi-view or
\clcl VLBI technique, which allows simultaneous
observations of a target and multiple calibrators around it
by replacing single telescopes by sites with multiple elements.
%The elements in a ``cluster'' are fed from a common 
%local oscillator and have independent pointing capabilities. 
%At each site, during the observations, individual or sub-groups of
%elements are allocated for the observations of each of the sources 
%(calibrators and target).
%Afterwards the data on the calibrators are combined to correct 
%the observations of the target source.
%
The suitability of this technique to address the
ionospheric effects has been demonstrated with joint observations 
between connected
interferometer arrays at 1.6 GHz of a target and three calibrator
sources (see Fig.~1). Despite these benefits its use has been limited by the
shortage of observing sites, and the complexity in its implementation.
%complexity involved in its implementation,
%[REMOVE? from the observation set-up (observing configuration), to the
%subsequent data analysis, including the processing at the
%correlator], 
We wish to revisit this technique now in the light of the next
generation of instruments for low frequency observations that will
become operational in the course of the next decade, for which the
multi-beam capability is an ``in-built'' feature, 
such as ASKAP in the near future, and SKA in the long term.
%[OR
%The multi beam capability is a novel and distint feature of these
%instruments and cluster cluster techniques offer the way to export it
%to VLBI observations. 
%OR
We believe that 
%In particular, We ``hypothesize'' that 
the implementation of \clcl techniques will enhance the 
performance of VLBI observations with ASKAP by providing higher precision 
astrometric measurements of many more targets, along with 
an extremely wide-field mapping capability.

\section {``Cluster-Cluster'' VLBI: Basics and Demonstration}
%/past results}

\emph {Cluster-Cluster} or multi-view VLBI is a technique that
replaces single-beam telescopes by sites with  
multiple-beam capabilities (a ``\emph{cluster}'')  \cite{sas91}. 
The \clcl concept offers the prospect of correcting
for phase perturbations arising from the spatial and temporal structure in the
troposphere and ionosphere, since such structure can be modelled using
simultaneous observations of multiple reference sources. 
%This can be achieved replacing the single telescopes by sites with multiple elements (a “cluster”) fed from a common local oscillator. 
%During the observations, the antennas of a
%cluster are pointed in different directions.
The basic requirements for a site to conduct \clcl
observations are that the elements (usually antennas) are fed from
a common frequency standard, be equipped with a precise phase
calibration system and have the ability for independent pointing.
This ability makes \emph{cluster} sites flexible 
to adapt the sensitivity of the observations, 
%for each source, 
by adapting the number of antennas allocated to the
observations of each source.\\
%(ability to point in 
%different directions and apply fringe stopping and delay control).

%FOR LATER:
%- Each site has to have multiple single-beam antennas, and/or
%multi-beam anten%na/e (for example, Parkes-64m

%with multi-beam receiver plus 12-m) 
%-------  

%At a cluster site, each element is pointed to a different source. This mode provides simultaneous observations of multiple sources irrespective of their angular separation. We call this mode Cluster-Cluster VLBI.

The feasibility and advantages of this approach have been demonstrated
with observations, for example, carried out in November 1999 with 3
interferometers, the VLA, 
%(in New Mexico, USA), 
WSRT 
%(in The Netherlands, EU) 
and MERLIN, 
%(in UK, EU), 
at 1.6 GHz 
%(see \cite{rio95},\cite{rio97},\cite{rio02},\cite{por02}
(see [3-6]
%Rioja \etal \,1995,1997,2002; Porcas \& Rioja 2002 
for a detailed description of all our observations). Each interferometer
observed four sources simultaneously, selected from the VLBA
calibrator list such that one of them, the target, lies in the sky
surrounded by the other three sources, the calibrators.  The angular
separations from the central target source were 1, 3 and 6 degrees,
as shown in Figure 1.
%Figure 1 shows their distribution in the sky and the angular separations
%to the central source, the target, 
%OR MAYBE ALL THIS AS FIGURE CAPTION?
%The sources were selected to meet the sky distibution criteria  according to th%e criteria to be strong and compact and with one of
%them surrounded by the other three.
%The sources are strong and compact with one of them
%surrounded by the other three.
%
%and with three of
%them san spatial
%distribution in the sky such that 3 of them would surround the
%position 
%spatial distribution 
%
%selected sources 
%strong and compact sources from the VLBA calibrator list, 
%with three of them surrounding the other.
%which with a spatial distribution 
%
%The spatial distribution of the
%sources in the sky is such that 3 of them, the calibrator sources, 
%surround the 4th source, the target, located in a central
%position. The angular separations to the target source are 
%$1^o$, $3^o$ and $6^o$,  as shown in Figure 1.
We carried out a comparative analysis of techniques which aimed to correct the
ionospheric distortion of the wavefront, that is, using a combination 
of the observations on the three calibrators, and using 
a single calibrator at a time. 
The combined analysis showed that a simple 2-D
linear interpolation of the ionospheric paths would
be sufficient to correct the phase of the target source at the central
position, using the phases measured on the three other sources.  
Moreover, the 2-D linear model derived 
from the simultaneous observations
of the three sources resulted in a superior phase compensation
compared to that from the
(phase) difference between each of the pairs to the central source, as done in
conventional phase referencing.
%, even for the pair which is only 1 deg away.
%for any of the 3 pairs, even for only
%1 degree away.  \\
Thus, our results demonstrated the superiority of the \clcl
technique to compensate the ionospheric errors in VLBI observations,
even with respect to a single calibrator only 1 degree away, at 1.6
GHz.

\section{ASKAP in the context of  ``Cluster-Cluster''  VLBI}

%FACTS: ASKAP description as an interferometer
ASKAP is a next generation mid-to-low-frequency 
radio interferometer and one of the SKA demonstration telescopes,
located in Western Australia.
It comprises 36 12-m multi-beam antennas
distributed over 6 km, with most of them lying within a
circle 2 km in diameter. 
%, (using phased array feeds,) which will provide 
%extremely wide fields of view of about 30 square degrees (with about 
%30 antenna beams). 
It incorporates novel receiver technology consisting of
phased-array-feeds and beam forming modules, which will result in
an extremely wide FoV of about 30 square degrees
in the observing frequency band from 0.7 to 1.8 GHz,
using about 30 beams per antenna. 
It will observe an instantaneous bandwidth equal to 300 MHz and 
2 polarizations. \\
%will provide extremely wide fields of view of about 30 square degrees
%The maximum distance bewteen antennas is 6 km, with most of them lying within a
%circle of 2 km in diameter. 
%ASKAP will be located in the Murchison Radio-astronomy Observatory in
%the Murchison Shire of Western Australia. 
%(ASKAP is located on the Murchison Radio-astronomy Observatory in the
%Murchison Shire of Western Australia.), about ... km Nort-east of Perth.

%FACTS: ASKAP description as a VLBI element
%The science program of ASKAP incorporates a high resolution component
ASKAP will participate in VLBI observations as a tied-array, 
%with subarraying capability, 
and will retain the unique multi-beam feature. The number of
planned tied-array beams will be at least equal to that of antenna
beams, and fully steerable within the wide antenna FoV.
In other words, ASKAP operating in VLBI mode will be equivalent 
%as a VLBI site will operate 
to having about 30 single-beam antennas, each with a
collecting area equivalent 
%up 
to a 64-m dish, and independent pointing within the wide ASKAP FoV 
of about 30 square degrees.
Hence, joint VLBI observations with a network of conventional 
single-beam antennas, with typical FoVs of about 0.5 to 1 
degree, would only make use of a small fraction of the capabilities of ASKAP.\\

%,in particular with the Australian VLBI network (LBA) using e-VLBI. 

%characteristics
%(Will have a smaller tied array beam) keeping the multi beam performance. 
%about one tied array beam per antenna beam, and fully steerable
%Goal is to have Minimum, One tied array beam per antenna beam. The
%tied array beams are fully steerable. 

%FACTS: ASKAP description as a cluster-cluster site
ASKAP 
%as a VLBI site (OR as a tied array)
%as a multi-beam site, (with its multi-beam) 
meets the criteria to be a \emph{cluster}, with the steerable multiple
tied-array beams taking the place of the steerable single-beam antennas.
To implement VLBI \clcl techniques
% (in cluster-cluster
%mode); OR implementing cluster-cluster techniques. \\
%**IN ??? \\
%The multi-beam operations == It is equivalent to having about 30 single beam an%tennas/elements 
%with a collecting area equivalent to a 64-m diameter dish each, 
%and independing pointing within about 30 square degrees area/region
%in the sky. The tied array beams have a size of about 10''.
%(INCLUDE ``in-beam'', very narrow tied array beam, vs cl-cl in
%discussions?
%and ALSO cl-cl as the carrier of multi beam to VLBI domain)
%Given its sensititivity would be a crutial element for ...
%
%FACTS: ASKAP partners for VLBI obs. 
%cl-cl observations - MAYBE move to DISCUSSION?
% VLBI observations with
%australian vlbi network (LBA) and other antennas with common visibility.
it is vital that the multi-beam operations of ASKAP
can be matched with similar multi-view capabilities from other
instruments which have common visibility with ASKAP. This is a list 
of candidate VLBI sites
that already support or can be upgraded to support multi-view VLBI
located in Australia and overseas. In Australia, 
%This is a list of potential partners that already support or can be 
%upgraded to support multi-view VLBI  in Australia are, the
the Australia Telescope Compact Array (ATCA), with 6 22-m antennas,
the Parkes Observatory, with a 64-m antenna and a 12-m antenna, both
of which will be equipped with multi-beam receivers, 
%12-m antenna testbed for phased-array feed prototype,
and the Mount Pleasant Radio Observatory, with three independent
antennas;  plus overseas, 
the Giant Meter-wave Radio Telescope (GMRT), in India,
and MeerKAT, South Africa's SKA pathfinder (see Fig.~2). \\
We are working towards testing the feasibility of
VLBI observations with a \clcl network of Australian
sites. Such a network would provide the necessary counterpart
to exploit the full capabilities of ASKAP in the VLBI domain.

\section{Discussion}

%Argument: Feasibility of cl-cl observations with ASKAP and LBA
%Conclusion (1): VLBI with ASKAP (and other LBA) in CL-CL: increased 
%astrometric performance: precision and scope of application
%Conclusion (2): VLBI with ASKAP (and other LBA) in CL-CL: wide FoV for mapping 
ASKAP fulfills the conditions to be a \emph{cluster}
site, and a key element for \clcl VLBI observations due
to its high sensitivity, 
which results from the large collecting area and wideband recording, and
the wide FoV.
%{\bf MORE CONCRETE --> correct ionosphere means higher astrometric precision}
The results of our \clcl experiment, with observations 
of a target surrounded by a group of three calibrators distributed across 
a region equivalent to the wide ASKAP FoV,
demonstrated the advantages of this configuration to
calibrate the ionospheric distortions, which degrade the positional accuracy
in conventional ``nodding'' VLBI observations at low frequencies. 
%/to match the spatial and temporal properties of the ionospheric dist
%and the suitability of any calibrator in the wide ASKAP field of view
%to calibrate such distortions at the position of the target source.
%showed that, as far as calibrators is concerned, any source within the
%wide ASKAP field of view can contribute to the calibration at the
%target position.
%to calibrate the ionosphere
%ionospheric modelling at a target position.
%suitability of any calibrator within the ASKAP field of
%view as a reference source . 
%as far as calibrators is concernend, 
ASKAP will be located in Western Australia, which happens to be a part 
of the world with common visibility with other VLBI sites which hold 
the potential to carry out multi-view VLBI.  
There are plans to study the needs to upgrade these
candidate sites to be \emph{clusters}  
and the feasibility of joint observations with ASKAP.
Three of the sites are members of
the Australian VLBI network (LBA) and routinely carry out joint observations.
Other candidate sites are located overseas, in India and Africa. 
In addition to those, any single-beam antenna site could become a
\emph{cluster} if equipped with a phased-array-feed receiver.
Therefore, we conclude that joint observations of ASKAP with other multi-view
sites using \clcl techniques would improve the outcomes of ASKAP 
%the high spatial resolution component of ASKAP 
applied to astrometric VLBI measurements, achieving higher precision
for many more, and a broader range of, targets, and with lower 
detection thresholds for imaging of weak sources.
% implementing cl-cl techniques in the 
%joint observations of ASKAP with other multi-view sites would 
%improve the outcomes h a cl-cl network, compared with
%single-beam conventional VLBI observations applied to astrometry, are:
%%[Target outcomes of such a cl-cl network are:] 
%improved astrometric
%accuracy, lower detection thresholds, and broader scope of feasible
%targets, since our demostration experiment shows that any calibrator
%in the wide FoV of ASKAP could be used for relative astrometry.
%
%implementation of cl-cl techniques 
%in joint observations of ASKAP with other multi-view sites would
%improve the outcomes of such a network achieving high precision
%astrometric measurements on a wider scope to many more targets, at low
%frequencies.
Also, \clcl techniques would export the extremely wide FoV
capability of ASKAP as a stand-alone instrument
%to the high spatial resolution component of ASKAP 
%to ASKAP in VLBI observations,
to VLBI observations with ASKAP, 
by enabling multiple wide-field VLBI mapping. \\

We discuss here the feasibility of VLBI astrometry at
low frequencies, based on the probability of finding suitable calibrators,
%using/for joint observations of ASKAP with 
using ASKAP either with a VLBI array of \emph{clusters} or with a single-beam
antenna network.
%a \clcl network and with a single-beam antenna network.
Best astrometry with a single-beam antenna network is carried out
using phase referencing techniques with an ``in-beam'' calibrator, or
the variation of this which is a weak ``in-beam'' calibrator
plus nodding. The
application of the ``in-beam'' technique, nevertheless, is constrained 
by the small probability of finding such a calibrator
% (even for the weak ``in-beam'' variation) 
within the same FoV, about 1 square degree, 
as the target.
% (i.e. < 1 degree away). 
For a given target, 
the probability would be higher the greater the sensitivity of the array.
%this probability increases with the sensitivity of
%the observing array.
On the other hand, the \clcl technique achieves 
comparable astrometric results with the additional advantage that it can
draw calibrators from a larger fraction of the sky, as wide as the
FoV of ASKAP, about 30 square degrees.
For comparison, we have done a rough estimate of the impact of the
increased sensitivity and wide FoV on the number of
astrometric calibrators, in observations of ASKAP with a
network of single-beam antennas and a network of \emph{clusters}, respectively.
The number of ``in-beam'' calibrator sources, required for astrometry
with ASKAP in a VLBI array of single-beam antennas,
will grow with the increased 
recorded bandwidth to the power of approximately 0.85, since the sensitivity is 
proportional to the square root of the bandwidth and assuming a 
logN-logS source distribution has an index 
of about $-$1.7. 
Hence a bandwidth increase by a factor of 10, which corresponds to 
an upgrade from the current LBA standard mode to the bandwidth of
ASKAP, would increase by about 7 times the number of ``in-beam'' calibrators.
In turn, the number of \clcl calibrator sources
for astrometry with ASKAP in a network of multi-view sites 
%using  \clcl techniques, 
will scale with the FoV, which is about 30 times more.
%A rough calculation showes that for a given target, the increase 
%in the number of calibrator sources 
%expected from the expanded ASKAP FoV, of interest for
%"Cluster-Cluster", is significantly larger than those from the
%increased sensitivity derived from wideband recording with ASKAP, 
%relevant for cl-cl and "in-beam" techniques, respectively. \\
The conclusion is that there are more sources to be had from an expanded 
FoV than from an increased bandwidth. 
Therefore, from the perspective of available calibrators, 
%the scope of the ``cl-cl'' would be larger than ``in-beam'' phase
%referencing technique/Therefore 
the \clcl technique has a more comprehensive application.
The two techniques are not exclusive, and their combined use would lead to the 
optimal outcome, which can be judged on a case-by-case basis. \\

%Argument: Thinking ahead, General use? Exclusive pathfinder with VLBI comp.
%Conclusion: CL-CL VLBI with other next generation instruments
%possible? Demonstration of high resolution component for SKA - 
Looking to the future, the ASKAP is a pathfinder for the SKA Program,
and one of the next generation telescopes that will see light in 
the coming decade. 
The outlined benefits of using \clcl techniques in VLBI observations with ASKAP 
would apply equally to some of those instruments.   
Ultimately, establishing a \clcl network for 
VLBI observations with ASKAP would contribute 
to developing new techniques that are relevant for future high
resolution observations with the SKA in the mid-frequency range, and
exploit the technological development for the SKA to improve 
%into a positive  impact on 
the scientific capabilities of VLBI.

%\vspace*{-5cm} 

\begin{figure}
  \begin{center}
    \includegraphics[width=6cm]{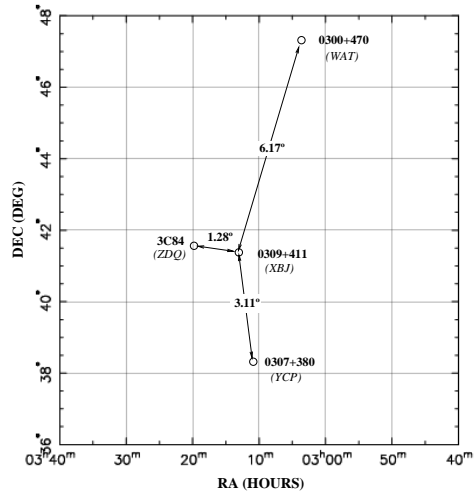}
  \end{center}
  \caption{Sky distribution of the sources simultaneously observed 
    in our \clcl observations carried out in 1999 with VLA, WSRT and
    MERLIN interferometers, at 1.6 GHz. The sources, strong and
    compact, belong to the VLBA calibrator list. The idea behind this
    technique is that the observations of a target source, in the
    middle, are calibrated using a combination of the phase
    measurements on the other three sources.}
\end{figure}

\begin{figure}
  \begin{center}
    \includegraphics[width=12cm]{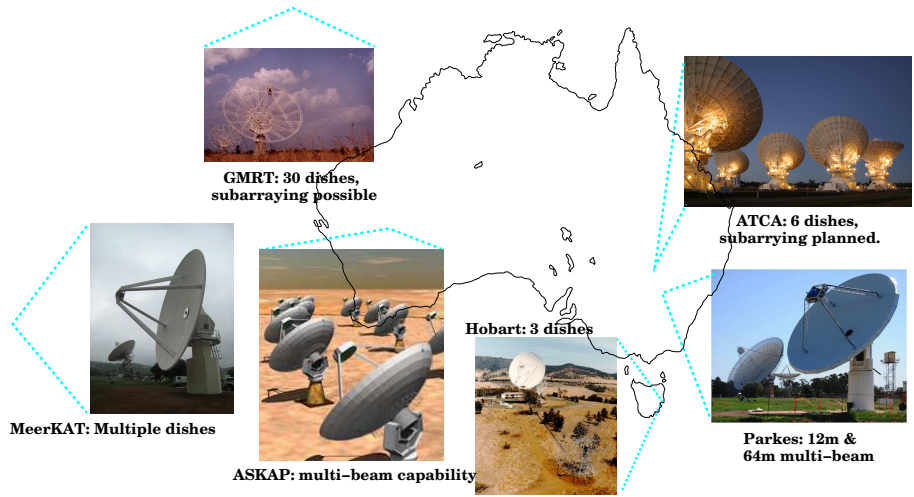}
  \end{center}
  \caption{Map of potential multi-view facilities in
    Australia and overseas for joint VLBI \clcl 
    observations with ASKAP. }
\end{figure}

\end{document}